\documentclass{article}
\usepackage[utf8]{inputenc}
\usepackage[rightcaption]{sidecap}
\usepackage{tikz}
\usepackage{authblk}
\usepackage[english]{babel}
\usepackage{hyperref}
\usepackage{ragged2e}
\usepackage[margin=0.8in]{geometry}
\usepackage{float}
\usepackage{xcolor}
\usepackage{fancyhdr}
\DeclareSymbolFont{matha}{OML}{txmi}{m}{it}
\DeclareMathSymbol{\varv}{\mathord}{matha}{118}
\usetikzlibrary{shapes.geometric, arrows}
\tikzstyle{used} = [rectangle, rounded corners, minimum width=2cm, minimum height=1cm, draw=green!60, fill=green!5, very thick, text centered]
\tikzstyle{parameter} = [rectangle, rounded corners, minimum width=2cm, minimum height=1cm, text centered, draw=blue!60, fill=blue!5, very thick,]
\tikzstyle{output} = [rectangle, rounded corners, minimum width=2.5cm, text width=2cm, minimum height=2cm, text centered, draw=yellow!60, fill=yellow!5, very thick,]
\tikzstyle{format} = [rectangle, rounded corners, minimum width=3.5cm,text width=3.5cm, minimum height=2.0cm, text centered, draw=red!60, fill=red!5, very thick,]
\tikzstyle{startstop} = [circle, minimum width=1.5cm, minimum height=1cm, text centered, draw=teal!60, fill=teal!5, very thick,]
\tikzstyle{arrow} = [thin,->,>=stealth]
\title{\textbf{Referencing Sources of Molecular Spectroscopic Data in the Era of Data Science: Application to the HITRAN and AMBDAS Databases}}
\definecolor{lime}{HTML}{A6CE39}
\DeclareRobustCommand{\orcidicon}{
	\begin{tikzpicture}
	\draw[lime, fill=lime] (0,0) 
	circle [radius=0.16] 
	node[white] {{\fontfamily{qag}\selectfont \tiny ID}};
	\draw[white, fill=white] (-0.0625,0.095) 
	circle [radius=0.007];
	\end{tikzpicture}
	\hspace{-2mm}
}
\foreach \x in {A,B,C,D,E,F}{\expandafter\xdef\csname orcid\x\endcsname{\noexpand\href{https://orcid.org/\csname orcidauthor\x\endcsname
			{\noexpand\orcidicon}}}}

\author[1,2,*]{Frances M. Skinner}
\author[1]{Iouli E. Gordon}
\author[4]{Christian Hill}
\author[1]{Robert J. Hargreaves}
\author[3]{Kelly E. Lockhart}
\author[1]{Laurence S. Rothman}
\affil[1]{Atomic and Molecular Physics, Center for Astrophysics| Harvard \& Smithsonian, Cambridge, MA USA}
\affil[2]{University of Massachusetts Lowell, Lowell, MA, USA}
\affil[3]{The SAO/NASA Astrophysics Data System (ADS), Center for Astrophysics|Harvard \& Smithsonian, Cambridge, MA, USA}
\affil[4]{Nuclear Data Section, International Atomic Energy Agency, Vienna International Centre, Vienna, Austria}
\affil[*]{Corresponding author E-mail address: frances.skinner@cfa.harvard.edu}

\date{\vspace{-4mm}\small\today}
\begin{document}
\maketitle
\vspace{-6mm}
\renewenvironment{abstract}
 {\small
  \begin{center}
  \bfseries \abstractname\vspace{-.5em}\vspace{0pt}
  \end{center}
  \list{}{%
    \setlength{\leftmargin}{0.5mm}
    \setlength{\rightmargin}{\leftmargin}
}
  \item\relax}
 {\endlist}
 \vspace{4mm}
\begin{abstract}
   The application described has been designed to create bibliographic entries in large databases with diverse sources automatically, which reduces both the frequency of mistakes and the workload for the administrators. This new system uniquely identifies each reference from its digital object identifier (DOI) and retrieves the corresponding bibliographic information from any of several online services, including the SAO/NASA Astrophysics Data Systems (ADS) and CrossRef APIs. Once parsed into a relational database, the software is able to produce bibliographies in any of several formats, including HTML and BibTeX, for use on websites or printed articles. The application is provided free-of-charge for general use by any scientific database. The power of this application is demonstrated when used to populate reference data for the HITRAN and AMBDAS databases as test cases. HITRAN contains data that is provided by researchers and collaborators throughout the spectroscopic community. These contributors are accredited for their contributions through the bibliography produced alongside the data returned by an online search in HITRAN. Prior to the work presented here, HITRAN and AMBDAS created these bibliographies manually, which is a tedious, time-consuming and error-prone process. The complete code for the new referencing system can be found at \url{https://github.com/hitranonline/refs}.
\end{abstract}
\section*{\small Introduction}\vspace{-2.5mm}\hrule\vspace{3mm}
Knowledge of spectroscopic parameters for transitions between energy levels in atoms and molecules is essential for interpreting and modeling the interaction of radiation (light) with different media. In order to aid researchers, spectroscopic parameters are being compiled into reference databases. In particular, the HITRAN (high-resolution transmission) molecular spectroscopic database\cite{gordon2017hitran2016} is a compilation of spectroscopic parameters used to simulate and analyze the transmission and emission of light in gaseous media, with an emphasis on planetary atmospheres. For half a century HITRAN has been considered to be an international standard which provides one recommended value per parameter for millions of transitions for different molecules.

HITRAN employs both experimental and theoretical data which are gathered from articles, books, proceedings, databases, theses, reports, presentations, unpublished data, papers in-preparation and private communications. Commencing with the HITRAN1986 edition\cite{HITRAN1986}, HITRAN started to provide reference mapping for the line positions, transition intensities, and broadening coefficients due to the pressure of air. From the HITRAN2004 edition\cite{HITRAN2004}, the majority of parameters in HITRAN have complete reference attributions. The current edition of HITRAN\cite{gordon2017hitran2016} contains references for dozens of parameters. It is imperative that all of these contributions to HITRAN receive acknowledgement through proper referencing to their cited material in the HITRAN database. This gives users an option to read more about how the parameters were determined and also acknowledges contributing papers and enables their citation by the users of the database. Furthermore, it assists the managers of the database in maintaining and updating complex segments of the database.

In this paper we describe a new, automated referencing system to provide consistent, accurate and detailed bibliographies to every source of data on the website. Starting from the HITRAN2016 edition\cite{gordon2017hitran2016}, the data is being distributed through HITRAN\emph{online} (\url{https://hitran.org}) which is built on a relational database model described in Hill et al.\cite{hill2016hitranonline}. The relational database approach removes the constraints of fixed-width text fields for the storage of parameters and allows an arbitrary number of parameters to be stored and retrieved for each transition, along with their uncertainties and bibliographic references.

Each data set returned by a search is accompanied by a bibliography. Each data file provides citations, hyperlinks and notes to the original data sources to make it easier for users to credit data providers. Bibliographies can be exported in several formats, including HTML, plain text and BibTeX. All of the information on the contributions to HITRAN has previously been entered manually into the database; this method is error-prone and time-consuming. The purpose of this work is to create a convenient bibliographic system, to enable contributors' work to be easily cited. Users utilizing this system need only enter a single line of information into the program, in order to obtain the complete bibliography entry for the paper they wish to cite. This system was designed to prevent common mistakes and ensure faster updates to the references system in HITRAN as well as to the Atomic and Molecular Bibliographic Data System (AMBDAS) database. We have previously reviewed existing data practices in molecular spectroscopy\cite{Gordon2016} and have identified that there is room for improvement which can be facilitated with specialized tools.  The goal of this work is to encourage an environment that promotes data sharing provenance and good practice amongst researchers and databases. 

The AMBDAS database is a collection of references to articles concerning collisional and spectroscopic processes in plasmas and plasma\textendash{}material interaction data, with a particular focus on their application to nuclear fusion energy research. The database website, accessible from \url{https://db-amdis.org/ambdas}, provides an interface for querying by collision species, process category and publication metadata. The bibliographic entries are maintained using the Python software described in this article, through established collaborations with the National Institute for Fusion Science (NIFS) in Japan, the National Fusion Research Institute (NFRI) in South Korea and the National Institute of Standards and Technology (NIST) in the United States of America, as well as through \emph{ad hoc} arrangements with individual consultants. The use of an intuitive, automated administration interface for importing data is an important way in which errors, ambiguities and duplications are minimized within the AMBDAS database.

The software tools presented here are described as applied to the HITRAN and AMBDAS databases, but are self-contained and can be applied to other database services. The complete code for the new referencing system is provided on an open-source and cost-free basis by Apache Software Foundation License\cite{Automatic_Referencing_System}, along with detailed instructions and resources for customizing the output formats of citations. The objective is that with enough availability and use, databases and researchers alike will be encouraged to share information between databases, in turn making their work more accessible to users. Enhancement of the HITRAN reference infrastructure will directly benefit the Virtual Atomic and Molecular Data Centre (VAMDC)\cite{Dubernet2016} and their recommended citation practices\cite{Zwolf2016,Zwolf2019}. 

This paper will not describe in detail the HITRAN data itself, searching methods, how to access data, graphing data nor other technical accessibility information. For details on this information, please refer to the papers describing the quadrennial HITRAN editions (the most recent one is HITRAN2016\cite{gordon2017hitran2016}) and Hill et al.\cite{hill2016hitranonline} article describing HITRAN$online$. The HITRAN\emph{online} website has been available at \url{https://hitran.org} since May 2015. Registration, at \url{https://hitran.org/register/} (and also linked from the home page), is free and requires the user to provide only a name and email address. 
\subsection*{\small Current Status of References in HITRAN}\vspace{-1.5mm}\hrule\vspace{3mm}
Changes to the referencing system in HITRAN is useful only if users are aware of how to access this information when using HITRAN's data access capabilities. Therefore this section is dedicated to providing a detailed review on how to retrieve bibliographies from HITRAN. There are several sections in HITRAN where the user may search for, graph, preview or download data. The corresponding bibliographies for these data are made available in several different ways which depend on the section the user is using at the time, as well as what specific information they are retrieving. 

First of all, in each bibliography the user will see a number assigned to every reference. This number is a unique ``global" identifying integer ID, which is referred to in relevant data files and is recorded for user accessibility and administrator storing purposes. Users of the legacy HITRAN 160-character .par format will find this "global" integer identification system convenient; the ``per-molecule" identifying integers are retained and cross-referenced with their global equivalents when this default output format (.par) is selected. Alternatively, a custom output format may be created and used as described in the article by Hill et al.\cite{hill2016hitranonline}.

\begin{figure}[htbp]
\centering
\includegraphics[width=11cm]{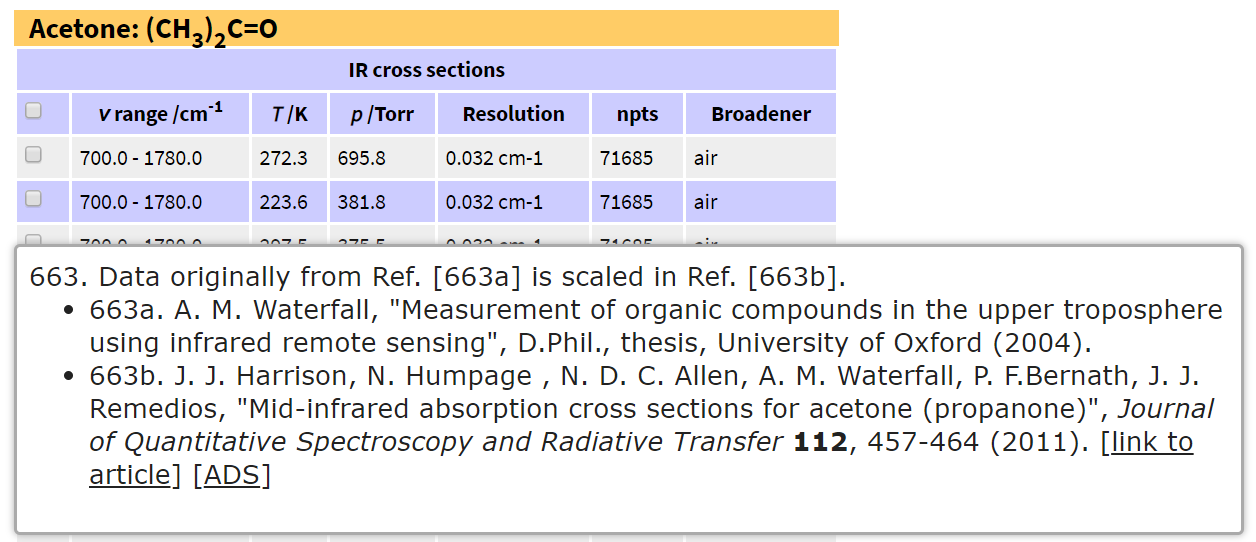}
\caption{The bibliography for a given temperature-pressure (T-p) absorption cross section of acetone. In this example, the cursor is over the second entry of the table. The number 663 to the left of the bibliography is the corresponding ``global" ID integer, and the two links at the end will take the user to the full-text article on the publisher's website or ADS, respectively. There are two references displayed in this bibliography for one line of data; these two are labeled separately 663a and 663b.}
\label{fig5}
\end{figure}

Some data in HITRAN are calculated from \emph{multiple} references and sources; to provide full credit to all contributors HITRAN nests multiple references under a single bibliographic entry. The main bibliography where multiple references are nested, is technically a ``note", while the nested references are complete bibliographies stored in the database. Therefore, any note can be created and assigned to data that is being referenced; the note will then pull the complete bibliographies of the papers that the data set is generated from. An example of this technique can be seen in Figure \ref{fig5}.

\begin{figure}[H]
\centering
\includegraphics[width=15 cm]{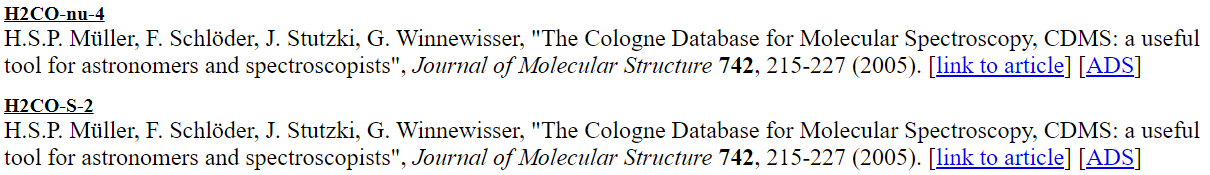}
\caption{An excerpt of the bibliography for the isotopologue H$_{2}^{12}$C$^{18}$O of formaldehyde in the line-by-line section of HITRAN\emph{online}. Every reference includes two hyperlinks at the end of each line that will take the user to the full-text article on the publisher's website or ADS, respectively.}
\label{fig1}
\end{figure}

\begin{figure}[H]
\centering
\includegraphics[width=13 cm]{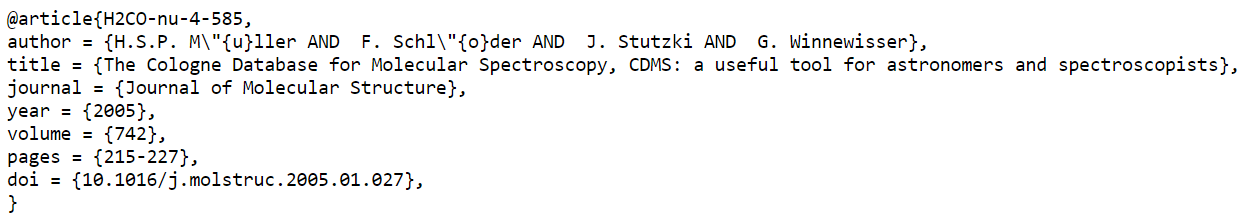}
\caption{An excerpt of the BibTeX formatted bibliography from HITRAN\emph{online} for use with \LaTeX. This bibliography is for the isotopologue H$_{2}^{12}$C$^{18}$O of formaldehyde in the line-by-line section.}
\label{fig2}
\end{figure}

\begin{figure}[H]
\centering
\includegraphics[width=14 cm, trim={0 0.7cm 0 0}, clip]{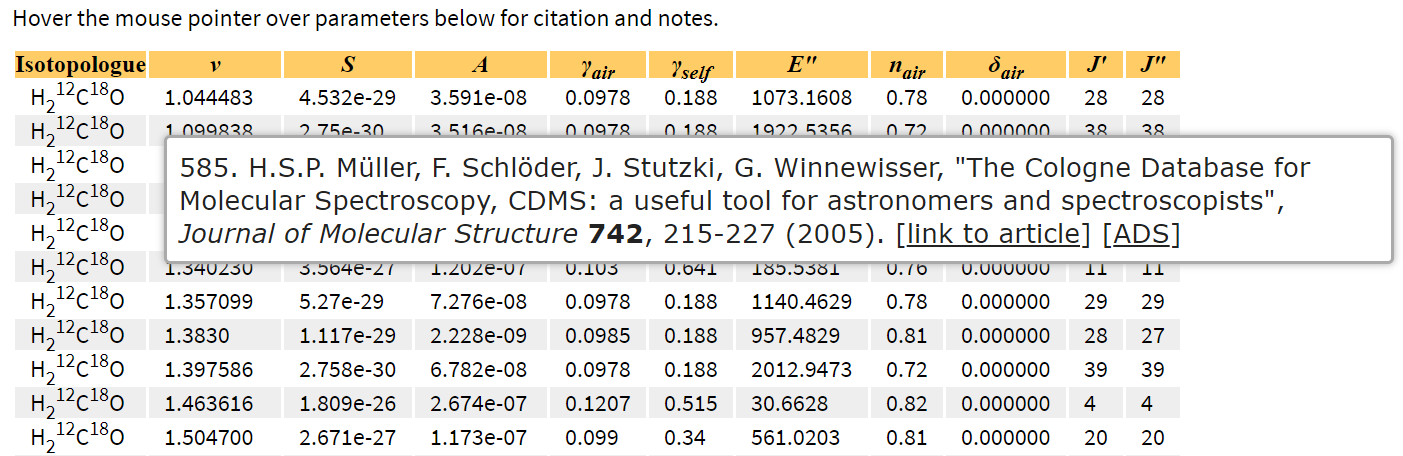}
\caption{A sample of ten transitions of the H$_{2}^{12}$C$^{18}$O isotopologue from the line-by-line section of HITRAN\emph{online}. This html table is displayed when there are fewer than 1\,000 transitions returned by the query.  A bibliography pop-up will appear for each parameter when the user hovers their cursor over the corresponding data. In this example, the cursor is over the wavenumber ($\tilde{nu}$) of the first row.}
\label{fig3}
\end{figure}

\begin{figure}[H]
\includegraphics[width=11 cm]{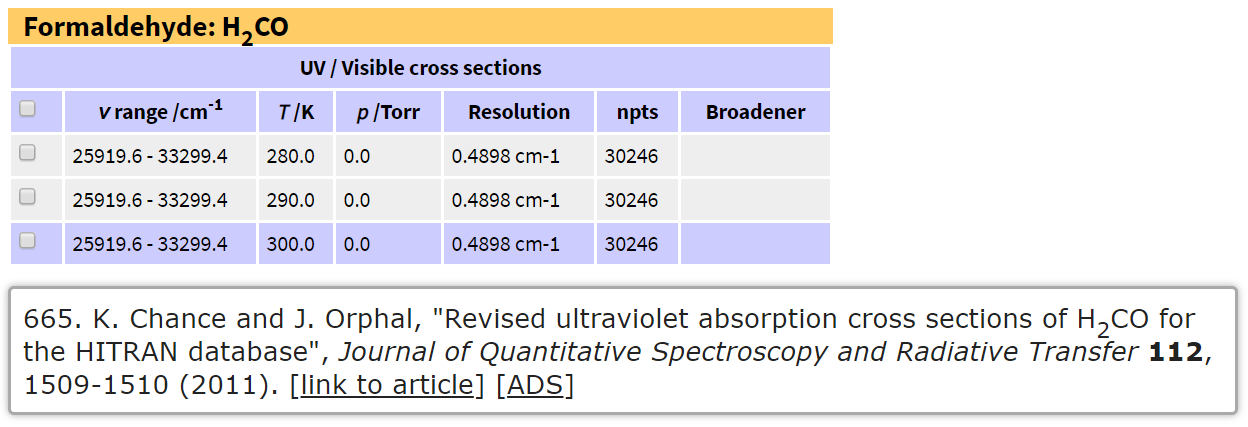}
\centering
\caption{The bibliography pop-up for a given temperature-pressure (T-p) absorption cross section of formaldehyde. In this example, the cursor is over the third entry of the table. The number 665 to the left of the bibliography is the corresponding ``global" ID integer, and the two hyperlinks at the end will take the user to the full-text article on the publisher's website or ADS, respectively.}
\label{fig4}
\end{figure}

HITRAN has several major sections that provide different types of spectroscopic data, with the traditional section being the line-by-line or molecular transition section. In the line-by-line section, when a user accesses their desired data, they will see a query-results web-page that will contain a``downloads" table. In this ``downloads" table, there are two bibliography files giving sources and notes relating to the returned data. One bibliography is an HTML file with links to the cited articles at their publisher websites and on the ADS database\cite{ADS} (Figure \ref{fig1}). The other bibliography file is a .bib file containing these references in BibTeX format that enables the inclusion into a LaTeX document (Figure \ref{fig2}). If there are fewer than 1,000 transitions returned by the query made by the user, then those transitions are listed in an HTML table on the same query results web page (Figure \ref{fig3}). Hovering the mouse cursor over each parameter in this table brings up a bibliography entry for that parameter which contains links to the article and any relevant notes on the reference.
In the absorption cross section part of HITRAN\emph{online} there is a complete list of references contained in the supplemental folder (provided at the top left of the window). In this folder the referenced sources can be found listed in HTML, Excel and plain text formats. If a user wishes to view the full bibliography for a particular absorption cross section, they click on their desired molecule and hover their mouse cursor over any of the rows of data listed for that particular molecule. An example is displayed for an absorption cross section of formaldehyde (Figure \ref{fig4}). After selecting one or more data sets, the user will be prompted to the final screen which then presents a page with links to the data files and a complete bibliography. 
In the collision induced absorption section of HITRAN\emph{online}, the corresponding references are given in the reference document PDF which is provided at the top of the web-page. In the aerosol properties section, the references for the utilized sources can be obtained through a PDF document. In the HITEMP section, users are asked to cite the original sources of data by using the assigned reference codes for each line transition, these reference codes are consistent with those in HITRAN line-by-line. In the supplemental section; there is a list of references at the bottom of the window and in each subsection (Line-Mixing, Total Internal Partition Sums, Supplemental Absorption Cross Sections and Radioactive Isotopologues); there exists an instruction manual containing the references used for the data provided. Overall these bibliographies are displayed so that users will understand where the data came from and to be able to use this information for their records and further research. HITRAN also provides the HITRAN Application Programming Interface (HAPI)\cite{HAPI} that allows downloading the spectroscopic data from HITRAN and carry out sophisticated calculations using predefined or custom functions. At the moment HAPI does not enable downloading reference data but the work is underway to provide this option in near future. 
\subsection*{\small Digital Object Identifier}\vspace{-1.9mm}\hrule\vspace{3mm}
A digital object identifier (DOI) is a string of numbers, letters and symbols used to permanently identify an article or document and link to it to its online original source; a DOI is assigned to almost every work that is published in the modern age\cite{IDF}. Even proceedings and people have their own DOI to electronically link public work to the sources. This DOI system is traceable and permanent, which is why HITRAN and AMBDAS chose to use the unique DOI for citing and referencing sources in their respective databases. A DOI is a unique text string assigned by the International DOI Foundation to identify content and provide a persistent link to its location on the internet. Digital objects may change physical locations, but the DOI assigned to that object will never change. Therefore by using the DOI the object is always accessible to the user. 

The DOI for an article can be found at the top or bottom corners of the published paper, or in the hyperlink to the paper. The new automated method implemented by the software described in this article enables the administrator to enter only the DOI for published work and in return retrieve the bibliography of the publication.
\section*{\small Results: Automatic Referencing System}\vspace{-2.8mm}\hrule\vspace{3mm}
This section describes the process of querying and retrieving bibliographies and references with the new automated referencing system. All code for the automatic referencing system was written in the Python programming language through the web-based, interactive computing notebook environment, Jupyter Notebook and is also implemented in a Django application available at \url{https://github.com/hitranonline/refs}. Interested users are encouraged to test the referencing system on this Jupyter notebook, which provides detailed instructions. Users who are interested in implementing the code through the provided Django application can find further details in the \texttt{setup.py}, \texttt{settings.py} and \texttt{README} files respectively. 

Several libraries are referenced, along with the necessary packages which must be installed and imported by the users. The referencing system was developed for the ease of use of the administrators of the HITRAN and AMBDAS database, as well as for the reliability of adding accurate information for the users and contributors of HITRAN alike. The automatic referencing system provides three different output formats for every reference generated. Several formats are necessary so that users have multiple options when viewing and accessing the bibliographies in HITRAN. The format outputs for references are as follows:
\begin{itemize}
\item HTML
\item JSON (Text)
\item BibTeX
\end{itemize}

Every referenced material in HITRAN also has the option of adding a detailed note to the bibliography. This detailed note option is consistently used for describing where data was taken from in the article, what data was not used and any other information necessary for understanding and using the referenced information. This same note option is included in the new automatic referencing system in HITRAN. Hyperlinks are included for all references as well, so that the users will have access to the full-text of the paper as well as the ADS link to the paper. The ADS link is a hyperlink to the ADS database, which provides bibliographic information to a majority of astronomical researchers worldwide. ADS provides several unique systems; the user has the option to view author network visualizations, paper citations, paper downloads, access citations in multiple formats, and account users have the option to develop a private library for their records. Thus, HITRAN has endeavored to include links to the ADS database for articles as well as the DOI links to published work.

The new referencing system is detailed in six clear steps, and a visualization is provided in Figure \ref{fig6}. However, if the administrators are referencing a source that exists in the ADS database, then they only need to follow three easy steps in order to cite the source properly in all three output formats and include relevant notes. 

The six steps are explained in further detail in the \hyperref[methods]{methods} section and they are listed as follows:
\begin{enumerate}
\item	Retrieve the DOI of the paper that is being cited. Enter the DOI in the prompt provided. 
\item	This step will populate the corresponding ADS bibcode for the paper. If no bibcode is generated, skip to step 4
\item	Use the retrieved bibcode from step 2 for the \hyperref[ads]{ADS method}. The output is customizable, all output formats are possible.
\item Use the DOI to search in \hyperref[urllib]{\texttt{urllib} method}. The output will be the bibliography as a JSON output.
\item  The citation from the \hyperref[urllib]{\texttt{urllib} method} will be automatically formatted in HTML in the following section titled \hyperref[json]{JSON-to-HTML}.
\item The initial DOI is automatically populated in the final \hyperref[bibtex]{BibTeX method} to retrieve the BibTeX citation for the paper.
\end{enumerate}
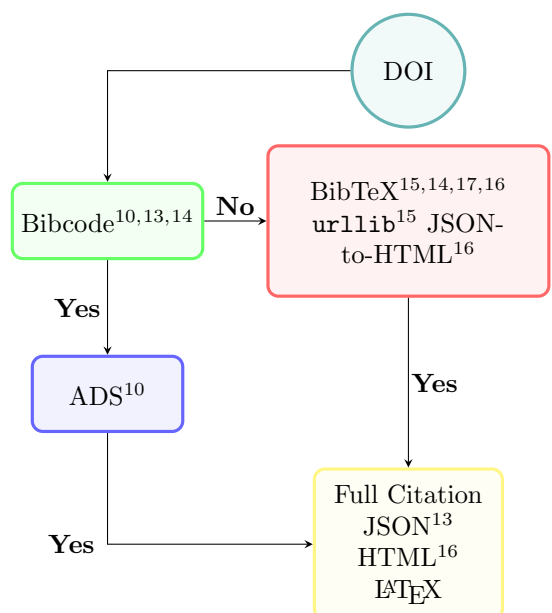
\begin{figure}[H]
\begin{center}
\begin{tikzpicture}[node distance=2cm]
\node (start) [startstop] {DOI};
\node (used) [used, below of=start, xshift=-4cm] {Bibcode\textsuperscript{10,}\textsuperscript{13,}\textsuperscript{14}};
\node (parameter) [parameter, below of=used, yshift=-0.3cm] {ADS\textsuperscript{10}};
\node (output) [output, below of=parameter, xshift=4cm] {Full Citation JSON\textsuperscript{13} HTML\textsuperscript{16} \LaTeX};
\node (format) [format, right of=used, xshift=2cm] {BibTeX\textsuperscript{15,}\textsuperscript{14,}\textsuperscript{17,}\textsuperscript{16} \texttt{urllib}\textsuperscript{15} JSON-to-HTML\textsuperscript{16}};
\draw [arrow] (start) -| (used); 
\draw [arrow] (used) -- node [xshift=-0.4cm] {\textbf{Yes}} (parameter);
\draw [arrow] (used) -- node [yshift=0.2cm] {\textbf{No}} (format);
\draw [arrow] (format) -- node [xshift=0.35cm] {\textbf{Yes}} (output);
\draw [arrow] (parameter) |- node [xshift=-0.48cm] {\textbf{Yes}} (output);
{\textbf{Yes}} (output);
\end{tikzpicture}
\caption{A diagram of the steps required to retrieve a reference from an article in multiple formats. The superscripts above the method names and outputs refer to the \hyperref[bibliography]{references} for those codes.}
\label{fig6}
\end{center}
\end{figure}
\vspace{-6mm}
Compared with manual entry, there is much less scope for human error in this new referencing method. The only part on the user side is to include any necessary notes required for citing the source properly. Users of the HITRAN database will see contributors' names and links to their publications on every window where their data is displayed, and all of this information will be accurate and up to date. This system of displaying contributors' work has been the conventional method used by HITRAN since 2015. Prior to this referencing system, there existed a static file containing all of the references in HITRAN at the time and to what molecules or data these references were referred to. 

Henceforth, HITRAN will continue the good-practices it has always used, for providing references of data shown in the database. This new automatic method for adding sources into HITRAN and AMBDAS will be the next stepping stone towards modernization of each database.
\section*{\small Discussion}\vspace{-1.9mm}\hrule\vspace{3mm}
HITRAN gets its data from contributors who publish their results, present their data at a conference or communicate privately with the HITRAN team. Adding these contributors' references in HITRAN is currently all done by hand. Therefore, this project created a new automated method for adding references in HITRAN, using only a DOI. This new automated system will be easier for the administrators/users to maintain, and data replacement will also be simplified as well as users' access to references. Most importantly, the contributors to HITRAN and AMBDAS will receive proper credit and representation. This change will ideally set an example to the research community and encourage a productive data sharing environment amongst researchers. 

Prior to creating the new automated method, references in the HITRAN database were updated manually, either by transcribing reference metadata or by cutting-and-pasting from websites, PDF files and so on. Every reference needed to be processed painstakingly by hand, including: correlating the bibliography information to the main article or resource, adding the DOI information on the article, checking spelling, adding separate plain text, \LaTeX~and HTML markups, etc. About fourteen hundred references were corrected with up-to-date information, and  their outputs were tested and double checked before updating the database. In addition, all changes are recorded for administrator record-keeping, and this further increases the time and attention required by the administrator. The new automated method of simply retrieving references with a single click of a button will ensure faster results, create a smaller time requirement and ensure minimal mistakes. 
\section*{\small Methods: Bibcode Method}\label{bibcode}\label{methods}\vspace{-1.5mm}\hrule\vspace{3mm}
In every method, including this method, several modules will need to be imported or installed. The modules that are used include \texttt{requests} and \texttt{JSON}. The \texttt{json} module of the Python standard library\cite{json} is necessary for obtaining JSON outputs from bibliography retrieval methods designed in the new automated referencing system. The \texttt{requests} module allows the user to send HTTP/1.1 requests, without the need for manually adding query strings to URLs, or to form-encode POST data\cite{requests}. 

In this method the administrator or user is required only to enter the DOI in the prompt provided. The user will then be shown the bibcode generated from the search through the ADS database. The bibcode shown will be populated throughout the entire \hyperref[ads]{ADS method} unless changed, allowing the user to only run their desired fields to retrieve custom formatted citations. The DOI entered in this initial prompt is also populated for the \hyperref[urllib]{\texttt{urllib} method} and the \hyperref[bibtex]{BibTeX method} for ease of use.

A bibcode is what the NASA Astrophysics Data System uses to identify literature in their database. The bibcode is a 19 digit identifier and takes the format YYYYJJJJJVVVVMPPPPA where:
\begin{itemize}
    \item YYYY: Year of publication
    \item JJJJJ: A standard abbreviation for the journal (e.g. ApJ, AJ, MNRAS, Sci, PASP, etc.). A list of abbreviations is available.
    \item VVVV: The volume number (for a serial) or an abbreviation that specifies what type of publication it is (e.g. conf for conference proceedings, meet for Meeting proceedings, book for a book, coll for colloquium proceedings, proc for any other type of proceedings).
    \item M: Qualifier for publication:
        \subitem E: Electronic Abstract (usually a counter, not a page number)
        \subitem L: Letter
        \subitem P: Pink page
    \item Q-Z: Unduplicating character for identical codes
    \item PPPP: Page number. Note that for page numbers greater than 9999, the page number is continued in the m column.
    \item A: The first letter of the last name of the first author.\footnote{NASA/ADS help page at \url{https://adsabs.github.io/help/actions/bibcode}, accessed on 4 September 2019.}
\end{itemize} 
The fields that are empty in a bibcode are replaced with periods (.) so that the code is always 19 characters long. As an example, the bibcode "2017JQSRT.203....3G " corresponds to the Journal of Quantitative Spectroscopy and Radiative Transfer (JQSRT), year 2017, volume 203, first page 3 and the first letter of the last name of the first author is G.

If a Bibcode was not generated in the Bibcode method, then that would mean the paper the administrator or user is looking for is not on the ADS database. This will require the user to skip the following \hyperref[ads]{ADS method} and move on to the \hyperref[urllib]{\texttt{urllib} method}, the \hyperref[json]{JSON-to-HTML} and the \hyperref[bibtex]{BibTeX method}. These three methods are fairly straight forward. The user will only need to run these coded cells for them to work properly since the necessary information is already automatically populated for the user, so no manual entering is required. However, these last three methods provide less customization options compared to the \hyperref[ads]{ADS method}. 
\subsection*{\small ADS Method}\label{ads}\vspace{-1.5mm}\hrule\vspace{3mm}
Use of the ADS API requires that the developer has read and agrees to the NASA/ADS terms and conditions which can be found at \url{https://adsabs.github.io/help/terms/}. In order to use the ADS method, in addition to the bibcode from the initial \hyperref[bibcode]{Bibcode method} step, the user will require a valid token in order to use ADS's API. A token can be issued when the user registers for an account on NASA/ADS at \url{https://ui.adsabs.harvard.edu/} and is required for every fetch of a bibliographic entry while using the ADS method. 
This method allows the user to customize the bibliography output in various ways. The set customization, already programmed for the HITRAN and AMBDAS administrators, is widely used. The structure of the reference output is as follows and in the following order: 
\begin{itemize}
    \item notes are entered first
    \item the authors are listed as F. I. Surname, with commas in between each author
    \item the title of paper is in quotations
    \item the journal is italicized
    \item the volume number is in bold
    \item the pages of the article are written as first page-last page
    \item the year of the article is written in parentheses
    \item hyperlinks are finally included at the end, one for the DOI hyperlink and one for the ADS hyperlink
\end{itemize}

ADS has made it available to include more details and information in bibliographies when using their database. A user has access to information such as: the abstract, copyright, citation count, author affiliation, keywords, publication category and the arXiv e-print number of the article, etc. The user also has more output options such as: EndNote, ProCite, RIS (Refman), RefWorks, MEDLARS, AASTeX, Icarus, MNRAS, Solar Physics (SoPh), DC (Dublin Core) XML, REF-XML, REFABS-XML, VOTables and RSS. See \url{https://adsabs.github.io/help/actions/export} for more details and information.

Most importantly, HITRAN and AMBDAS chose to utilize ADS because of efficiency. The administrator has the option to search more than one bibcode at a time, allowing for multiple reference retrievals in any of the three designated formats desired. All the administrator will need to do is enter their personal token, and their bibcode(s). There are three formats that HITRAN uses for its reference methods: BibTeX, JSON and HTML. Therefore there are three main search parameters for users to use when creating bibliographies, all three of these are already custom formatted.
\subsection*{\small urllib Method}\label{urllib}\vspace{-1.5mm}\hrule\vspace{3mm}
\texttt{urllib} is a package that collects several modules for working with URLs. The two sub-modules used in this method are \texttt{urllib.request} and \texttt{urllib.error}. The \texttt{urllib.request} module defines functions and classes which help in opening URLs, basic and digest authentication, re-directions, cookies and more. While the \texttt{urllib.error} contains the exceptions raised by \texttt{urllib.request}\cite{urllib}.

This method titled \texttt{urllib} method, provides a JSON formatted output along with hyperlinks and the source DOI. The administrator need only enter the DOI for the article or paper that is being cited into the program. Upon doing so and running the method, the user will retrieve the complete JSON bibliography of the paper. The administrator is not required to complete this method or the following methods titled \hyperref[json]{JSON-to-HTML} or \hyperref[bibtex]{BibTeX method}; the \hyperref[ads]{ADS method} will provide the desired outputs that would be generated in these methods. This method and those following are made available in case the paper is not in the ADS database only, since the ADS contains items within (or relating to) the field of astrophysics.
\subsection*{\small JSON-to-HTML Method}\label{json}\vspace{-1.5mm}\hrule\vspace{3mm}
In this method, the administrator will need to import the Python module \texttt{html}\cite{html} only. The JSON-to-HTML method provides an HTML encoded output for the administrator while the only information it is given is the JSON encoded bibliography from the \hyperref[urllib]{\texttt{urllib} method}. The entire JSON output from the \hyperref[urllib]{\texttt{urllib} method} will automatically be populated into the designated region in this method. Therefore, the administrator need only to run the method to retrieve the now HTML encoded bibliography. The html encoder will replace the following characters (\& < " ' >) with recognized html entities (\&amp; \&lt; \&quot; \&\#x27; \&gt;) respectively.
\subsection*{\small BibTeX Method}\label{bibtex}\vspace{-1.5mm}\hrule\vspace{3mm}
\linespread{1.4}{
In this method there are several modules and packages that the administrator will need to install or import, including: \texttt{urllib.request} with the sub-modules \texttt{quote}, \texttt{Request} and \texttt{urlopen}\cite{urllib}. Users will also need to install the \texttt{BeautifulSoup} package\cite{bs4}, and import the Python standard library packages \texttt{re} (for regular expression matching), \texttt{logging} (for applications to configure different log handlers and a way of routing log messages to these handlers) and the \texttt{html.entities} sub-module \texttt{name2codepoint}.
\newline
\indent
This particular method titled BibTeX method was designed so that the user's initial DOI entered in the \hyperref[bibcode]{Bibcode method} would automatically populate in this method. Once the administrator runs the code with the entered DOI, they will be shown a detailed bibliography, in \LaTeX \hspace{0.2mm} format, with labels to all the corresponding information retrieved. The bibliography will contain, from top down, the paper's title, list of authors, journal, volume, number, pages, year and publisher. The only requirement on the user's side of things, is to enter the DOI for the paper or article into the initial prompt in the \hyperref[bibcode]{Bibcode method} and run that program first. 

It is important to note that this method is the only one that will be able to also work for other keywords when searching for bibliographies. For example, if one were to search the title of the paper, instead of the paper's DOI, then they would receive a full bibliography corresponding to the information entered into the program. However, this bibliography may resolve to an incorrect article or paper. Therefore, even though this is a convenient option of simply entering the title of the paper, it will still be encouraged for users to use the DOI of the article instead.}
\section*{\small Conclusion}\vspace{-2.2mm}\hrule\vspace{3mm}
In this paper we describe a new, automated referencing system to provide consistent, accurate and detailed bibliographies to every source of data in scientific databases. The goal of this work is to encourage an environment that promotes data sharing provenance and good practice amongst researchers and databases. Overall, it is imperative that all contributions to scientific databases receive acknowledgement through proper referencing to their cited material. All code for the automatic referencing system was written in the Python programming language through the web-based, interactive computing notebook environment, Jupyter Notebook and is also implemented in a Django application available at \url{https://github.com/hitranonline/refs}.

This work provides a convenient bibliographic system, to allow database administrators to implement bibliographies faster and without human errors into their database systems. Users utilizing this system need only enter a single line of information into the program in order to obtain the complete bibliography entry for the paper they wish to cite. This system was designed and tested for the HITRAN and AMBDAS databases, but can also be used through the raw code outlined in the Jupyter Notebook or integrated into a database through the Django application. The complete code for the new referencing system is provided through a permissive, open-source and cost-free basis by Apache Software Foundation License\cite{Automatic_Referencing_System}, along with detailed instructions and resources for customizing the output formats of citations and implementation of the Django application. Ideally, with enough availability and use of proper referencing systems, databases and researchers alike will be encouraged to share information between databases, in turn making their work more accessible to users.
\vspace{2cm} 
\newline
\footnotesize
\linespread{1.0}
\section*{\footnotesize Author Contributions}\vspace{-2.0mm}\hrule\vspace{3mm}
{conceptualization, Iouli Gordon, Laurence Rothman; methodology, Iouli Gordon, Christian Hill, Frances Skinner; software, Frances Skinner, Robert Hargreaves, Kelly Lockhart and Christian Hill; validation, Iouli Gordon and Frances Skinner; formal analysis, Iouli Gordon; investigation, Frances Skinner, Christian Hill; resources, Iouli Gordon, Frances Skinner and Kelly Lockhart; data curation, Frances Skinner, Christian Hill; writing--original draft preparation, Frances Skinner; writing--review and editing, Iouli Gordon, Christian Hill; visualization, Frances Skinner; supervision, Iouli Gordon; project administration, Iouli Gordon; funding acquisition: Foundation under Grant No. 1745460.}
\section*{\footnotesize Funding}\vspace{-2.5mm}\hrule\vspace{3mm}
{FMS acknowledges the internship though the Smithsonian Astrophysical Observatory Latino Initiative Program funded by the National Science Foundation under Grant No. 1745460. The HITRAN database is supported through the NASA Aura and PDART grants NNX17AI78G and NNX16AG51G, respectively.}
\section*{\footnotesize Acknowledgments}\vspace{-2.5mm}\hrule\vspace{3mm}
{Joshua Karns for technical support while coding in Python. The NASA Astrophysics Data System allowed use of their API during development of this software and provided helpful personal guidance and assistance.}
\section*{\footnotesize Conflicts of Interest}\vspace{-2.0mm}\hrule\vspace{3mm}
{The authors declare no conflict of interest. The funders had no role in the design of the study; in the collection, analyses, or interpretation of data; in the writing of the manuscript, or in the decision to publish the results}
\vspace{-1.5mm}
\section*{\footnotesize Abbreviations}\vspace{-2.0mm}\hrule\vspace{3mm}
{The following abbreviations are used in this manuscript:\newline
\noindent 
\begin{tabular}{@{}ll}
ADS (NASA Astrophysics Data System)\\
AMBDAS (Atomic and Molecular Bibliographic Data System)\\
API (Application Programming Interface)\\
DOI (Digital Object Identifier)\\
HAPI (HITRAN Application Programming Interface)\\
HITEMP (High-Temperature Molecular Spectroscopic Database)\\
HITRAN (High-Resolution Transmission Molecular Absorption Database)\\
HTML (Hypertext Markup Language)\\
JSON (JavaScript Object Notation)\\
NASA (National Aeronautics and Space Administration)
\end{tabular}}
\medskip
\renewcommand{\refname}{\footnotesize Bibliography\vspace{0.5mm}\hrule\vspace{2mm}}
\bibliographystyle{unsrt}
\linespread{1.0}
\bibliography{main.bib}\label{bibliography}
\end{document}